\documentclass[twocolumn, 10pt, letterpaper]{article}
\usepackage[pass, margin=1in]{geometry} %one inch margins
\usepackage{ibpsa} % defines look
\usepackage{pslatex}
\usepackage{achicago}
\usepackage{amsmath, amssymb}
\usepackage{graphicx}
\usepackage{fancyhdr}
\usepackage{ifthen}
\usepackage[dvipsnames]{xcolor}
\usepackage{makecell}

\usepackage[acronym, nonumberlist]{glossaries}
\makeglossaries
\setacronymstyle{long-short}
\newacronym[longplural=cluster validity indices]{cvi}{CVI}{cluster validity index}
\newacronym{adr}{ADR}{automated demand response}
\newacronym{ai}{AI}{artificial intelligence}
\newacronym{amy}{AMY}{actual meteorological year}
\newacronym{bau}{BAU}{business-as-usual}
\newacronym{bes}{BES}{building energy system}
\newacronym{bess}{BESS}{battery energy storage system}
\newacronym{caidi}{CAIDI}{Customer Average Interruption Duration Index}
\newacronym{cdd}{CDD}{cooling degree days}
\newacronym{der}{DER}{distributed energy resource}
\newacronym{dhw}{DHW}{domestic hot water}
\newacronym{dp}{DP}{dynamic programming}
\newacronym{drl}{DRL}{deep reinforcement learning}
\newacronym{dr}{DR}{demand response}
\newacronym{dtw}{DTW}{Dynamic Time Warping}
\newacronym{ercot}{ERCOT}{Electric Reliability Council of Texas}
\newacronym{ess}{ESS}{energy storage system}
\newacronym{eulp}{EULP}{End-Use Load Profiles}
\newacronym{ev}{EV}{electric vehicle}
\newacronym{gan}{GAN}{generative adversial network}
\newacronym{gbdt}{GBDT}{gradient-boosted decision trees}
\newacronym{geb}{GEB}{grid-interactive efficient building}
\newacronym{ghg}{GHG}{greenhouse gas}
\newacronym{hdd}{HDD}{heating degree days}
\newacronym{hvac}{HVAC}{heating ventilation and air conditioning}
\newacronym{ieq}{IEQ}{indoor environmental quality}
\newacronym{kpi}{KPI}{key performance indicator}
\newacronym{lstm}{LSTM}{long short-term memory}
\newacronym{mdp}{MDP}{markov decision process}
\newacronym{mpc}{MPC}{model predictive control}
\newacronym{occ}{OCC}{occupant-centric control}
\newacronym{ppo}{PPO}{proximal policy optimization}
\newacronym{pv}{PV}{photovoltaic}
\newacronym{rbc}{RBC}{rule-based control}
\newacronym{res}{RES}{renewable energy source}
\newacronym{rlc}{RLC}{reinforcement learning control}
\newacronym{rlcc}{RLC}{resistor, inductor, and capacitor}
\newacronym{rl}{RL}{reinforcement learning}
\newacronym{rnn}{RNN}{recurrent neural network}
\newacronym{sac}{SAC}{soft actor-critic}
\newacronym{saifi}{SAIFI}{System Average Interruption Frequency Index}
\newacronym{soc}{SoC}{state-of-charge}
\newacronym{soo}{SOO}{sequence of operation}
\newacronym{tes}{TES}{thermal energy storage}
\newacronym{tou}{ToU}{time-of-use}
\newacronym{usa}{U.S.}{United States}
\newacronym{v2g}{V2G}{vehicle-to-grid}
\newacronym{vpp}{VPP}{virtual power plant}
\newacronym{zne}{ZNE}{zero-net energy}

\usepackage{subcaption}

\usepackage{algorithm}
\usepackage{algpseudocode}

\usepackage{bm}

\usepackage{amssymb}% http://ctan.org/pkg/amssymb
\usepackage{pifont}% http://ctan.org/pkg/pifont

\usepackage{tabularx}
\usepackage{{makecell}}
\setcellgapes{3pt}\makegapedcells
\usepackage{array,collcell}
\newcommand\AddLabel[1]{%
  \refstepcounter{equation}% increment equation counter
  (\theequation)% print equation number
  \label{#1}% give the equation a \label
}
\newcolumntype{M}{>{\hfil$\displaystyle}X<{$\hfil}} % mathematics column
\newcolumntype{L}{>{\collectcell\AddLabel}r<{\endcollectcell}}

\usepackage{varioref}
\usepackage[hidelinks]{hyperref}
\usepackage[noabbrev,capitalise]{cleveref}

\let\comment\undefined
\usepackage[final]{changes}
\definechangesauthor[name=del, color=red]{del}
\newcommand{\stkout}[1]{\ifmmode\text{\sout{\ensuremath{#1}}}\else\sout{#1}\fi}
\setdeletedmarkup{\stkout{#1}}

\renewcommand{\eqref}[1]{(\ref{#1})}

\begin{document}
\date{}

\fancypagestyle{empty}{%
  \fancyhf{}% Clear header/footer
  
  \vspace{0.5in}%
  \fancyhead[L]{
  \begin{tabular}{cc}
   \includegraphics [width=0.48in] {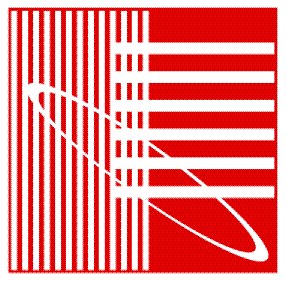} & \makecell[l]{\hspace{-0.6em}SimBuild \\ \hspace{-0.6em}2024 \vspace{.6em} \\ ~} \\
  \end{tabular}
  } % conference logo and title
  \fancyhead[R]{Eleventh National Conference of IBPSA-USA \\ Denver, Colorado \\ May 21 -- 23, 2024}% header text
}

\ifthenelse{\value{page}=1}{\setlength{\textheight}{8.5in}}{\setlength{\textheight}{9in}}

\title{\vspace{-0.4in} \titlefont % do not change this line
    Applications in CityLearn Gym Environment for Multi-Objective Control Benchmarking in Grid-Interactive Buildings and Districts\vspace{-0.2in} % do not change this line
}% do not change this line

\author{%
    \authorfont{~}\\% do not change this line
    \authorfont{
        Kingsley Nweye$^1$
        and Zoltan Nagy$^1$
    }\\
    \authorfont{$^1$The University of Texas at Austin, Austin, Texas, USA}
    % \authorfont{The names and affiliations SHOULD NOT be included in the draft submitted for review.}\\
    % \authorfont{The header consists of 10 lines with exactly 14 point spacing.}\\
    % \authorfont{The line numbers are for information only. The last line below should be left blank.}\\
    \authorfont{~}\\ % used to add blank lines
    \vspace{-0.55in} % do not change this line
}

\maketitle
\thispagestyle{empty}

% --------------------------------------------------------
\section{Abstract} \label{sec:abstract}
It is challenging to coordinate multiple \acrlongpl{der} in a single or multiple buildings to ensure efficient and flexible operation. Advanced control algorithms such as \acrlong{mpc} and \acrlong{rlc} provide solutions to this problem by effectively \deleted[id=del]{carrying out}\added{managing} a distribution of \acrlong{der} control tasks while adapting to unique building characteristics, and cooperating towards improving multi-objective \acrlong{kpi}. Yet, a \deleted[id=del]{major challenge}\added{research gap} for advanced control adoption is the ability to benchmark algorithm performance. CityLearn\deleted[id=del]{ is}\added{ addresses this gap} an open-source Gym environment for the easy implementation and benchmarking of simple \acrlong{rbc} and advanced algorithms that has an advantage of modeling simplicity, multi-agent control, district-level objectives, and control resiliency assessment. Here we demonstrate the functionalities of CityLearn using 17 different building control problems that have varying complexity with respect to the number of controllable \acrlongpl{der} in buildings, the simplicity of the control algorithm, the control objective, and district size.
\section{Introduction} \label{sec:introduction}
The electricity grid in the \gls{usa} is undergoing system-wide changes due to the electrification of buildings and transport systems\deleted[id=del]{, and increased unpredictability in electricity demand} with the goal of creating a robust and resilient electricity network and reducing carbon footprint \cite{us_department_of_energy_next-generation_2021}. Buildings are increasingly seen as participatory actors within the electricity market with the addition of on-site \glspl{res} and \glspl{ess}\deleted[id=del]{, as well as the shift from fossil-fueled end-uses to electric-powered end-uses such as heat pumps for space heating. Prediction of building energy demand has also increased in complexity, with energy-intensive appliances such as washing machines, dishwashers, and \glspl{ev} charging stations proliferating the residential sector. The adoption of work-from-home policies also brings about new challenges in the stochasticity of occupant behaviors and influence on energy demand patterns} \cite{neukomm_grid-interactive_2019}. \added{However, these \glspl{der} cause a deviation from the business-as-usual energy profile, introduce new peaks, and increase the complexity of building energy demand prediction. }At the urban scale, this dynamic and increasing building energy demand translates to periods of critical demand peak that threaten grid resiliency \cite{do_spatiotemporal_2023}. Furthermore, extreme-weather events such as heat waves\deleted[id=del]{,}\added{ and} winter storms\deleted[id=del]{, and hurricanes, which are increasing in frequency} due to climate change, \deleted[id=del]{can cause both increases in demand and decreases in supply due to}\added{exacerbate the risk of} power outages \cite{wei_energy_2023}.

\Glspl{der} including \glspl{res}, \glspl{ess}, and heat pumps that are installed at the point of consumption i.e., buildings, can provide the grid with energy flexibility in response to grid signals, changing weather conditions, or occupant preferences \cite{jensen_iea_2017}. Temporary load shedding, load shifting to off-peak hours, load modulation, and management of renewable power generation are several ways in which buildings can activate their flexibility \cite{neukomm_grid-interactive_2019}. It is, however, challenging to coordinate multiple \glspl{der} in a single or multiple buildings to ensure efficient and flexible operation. Advanced control algorithms such as \gls{mpc} \cite{drgona_all_2020} and \gls{rlc} \cite{nagy_ten_2023-1} provide solutions to this problem by effectively carrying out a distribution of \gls{der} control tasks while adapting to unique building characteristics, and cooperating towards improving multi-objective \glspl{kpi}.

A \deleted[id=del]{major challenge for}\added{research gap in the adoption of}  advanced control \deleted[id=del]{adoption }in \gls{dr} is the ability to benchmark algorithm performance in buildings \cite{vazquez-canteli_reinforcement_2019}. Physics-based models accurately capture the thermodynamics in buildings and are the industry standard for assessing the impact of control in the built environment\added{.} \deleted[id=del]{h}\added{H}owever, they require domain knowledge and attention to detail to model \glspl{bes} as well as significant effort to integrate standardized advanced control libraries in order to co-simulate and benchmark control algorithms. Building emulators provide various levels of abstractions of \glspl{bes}, which enables the designer to focus on the control implementation. Such emulators include Energym \cite{scharnhorst_energym_2021}, and BOPTEST \cite{blum_building_2021}, which provides high-fidelity energy models for control algorithm benchmarking. Although these emulators take advantage of robust simulation engines such as EnergyPlus and Modelica, their dependency on such engines raises the entry level for users. Moreover, they are designed for specific system-level or building-level environments and do not allow for district or neighborhood-level control or objectives. The more recent DOPTEST \cite{arroyo_prototyping_2023} framework builds upon BOPTEST to support district-level and multi-agent control but requires compute-intensive co-simulation and is limited to building and district environments provided within the framework.

CityLearn is an open-source Gym environment for the easy implementation and benchmarking of \gls{rbc}, \gls{rlc} and \gls{mpc} algorithms in a demand response setting \cite{nweye_citylearn_2023}. CityLearn is used to reshape the aggregated curve of electricity demand by controlling the energy storage of a diverse set of buildings in a district, thus allows for multi-agent control and district-level objectives. The initial release of CityLearn was designed to only manage \gls{ess} control for load shifting without altering indoor temperature conditions by making use of pre-defined building thermal loads \cite{vazquez-canteli_citylearn_2019}. Thus, the environment excluded a temperature dynamics model since ideal loads were maintained. This approach, however, left out the flexibility potential of electric-source \gls{hvac} systems e.g., heat pumps to provide load shedding energy flexibility, preheating, and pre-cooling to improve energy efficiency and maintain comfort. The latest CityLearn v2 release, at the time of writing, incorporates a temperature dynamics model that allows for heat pump power control to provide partial load satisfaction.

\added{Thus, our work here addresses the aforementioned research gap of benchmarking algorithm performance in buildings by providing environment and control examples as well as their implementation source code for members of the building control community whom are interested in using the latest CityLearn version to benchmark control algorithms for \gls{bes} management}. \deleted[id=del]{Thus, the objective of our work is to}\added{We} demonstrate different control tasks of differing complexity that can be tackled in CityLearn v2 where complexity refers to (1) the number of controllable \glspl{der} present in a building including \gls{dhw} \gls{tes}, \gls{bess} when paired with \gls{pv} system for self-generation and heat pump, (2) the simplicity of the control algorithm i.e., explainable \gls{rbc} or adaptive but black-box \gls{rlc}, (3) the control objective including reduction in electricity consumption, cost, \gls{ghg} emission, discomfort, peak demand, and their combinations, and (4) the size of the district i.e., number of buildings. \deleted[id=del]{Our contribution is to provide environment and control examples as well as their implementation source code for members of the building control community whom are interested in using CityLearn to solve their \gls{bes} control problems. }While the provided examples make use of \gls{rbc} and \gls{rlc} algorithms, these can be substituted with other control algorithms such as \gls{mpc}. The remainder of our paper is structured as follows: we provide details about CityLearn's functionalities and how it interfaces with control agents in the \nameref{sec:citylearn} section and describe our control simulation examples as well as the dataset used in these examples in the \nameref{sec:simulation} section. The results of the simulations are presented in the \nameref{sec:results} section and discussed in the \nameref{sec:discussion} section. Finally, we summarize our findings and set the stage for future work in the \nameref{sec:conclusion} section.
\section{CityLearn} \label{sec:citylearn}
\begin{figure}[!htb]
    \centering
    \includegraphics[width=0.90\columnwidth]{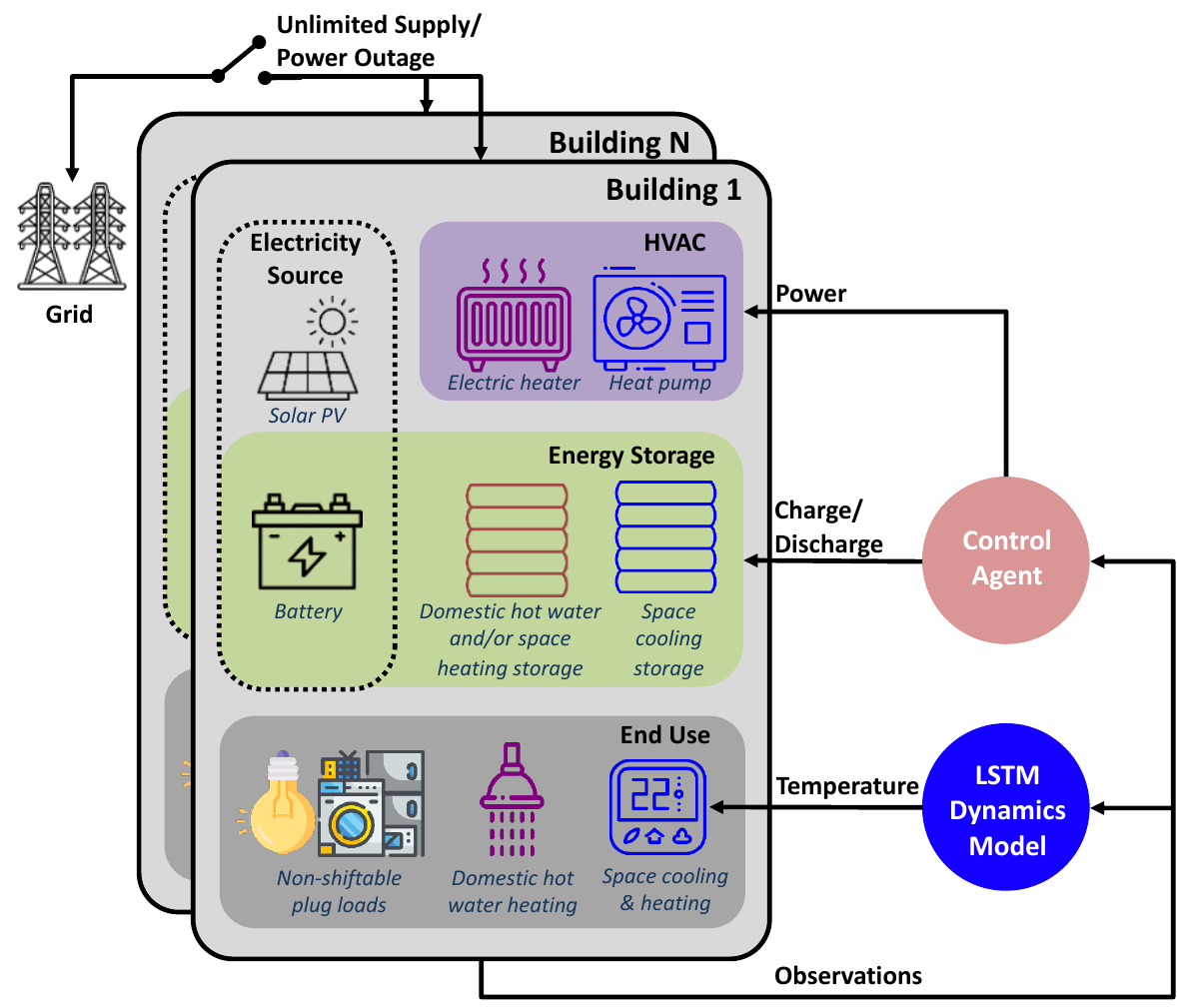}
    \caption{CityLearn environment and control interaction.}
    \label{fig:citylearn_systems}
\end{figure}

The CityLearn environment (\cref{fig:citylearn_systems}) consists of simplified building energy models that contain \gls{hvac} systems (heat pumps and electric heaters) and optional \gls{bess} and \gls{tes}. \added{The dynamics of these energy systems are based on reduced-order thermodynamic models that are documented in }\cite{vazquez-canteli_citylearn_2019}\added{.} Each building's space cooling, space heating and \gls{dhw} heating loads are independently satisfied through air-to-water heat pumps. Alternatively, heat pumps may be replaced with electric heaters to satisfy space and \gls{dhw} heating loads. \Glspl{tes} are charged by the \gls{hvac} system that satisfies the end-use, which the stored energy services. All \gls{hvac} systems as well as plug loads consume electricity from any of the available electricity sources including the grid, \gls{pv} system, and \gls{bess}. \Glspl{rbc}, \gls{rlc} or \gls{mpc} agent(s) are then used to manage load shifting in the buildings by determining how much energy to store or release at each control time step. The control architecture is either one agent to many buildings (centralized) or one agent to one building (independent) with optional information sharing amongst agents to achieve cooperative or competitive objectives. 

The agents can also, control the available power from the \gls{hvac} system to shed space thermal loads, preheat, or pre-cool the buildings. The consequence of the controlled power on the indoor dry-bulb temperature, i.e., the building dynamics, is modeled using a \gls{lstm} surrogate model based on the methodology by Pinto et al. \cite{pinto_data-driven_2021}. This methodology leverages EnergyPlus to generate training data including cooling or heating load and indoor temperature observations by varying the proportion of cooling or heating ideal load satisfied in conditioned zones in a building. These observations are combined with temporal and weather variables for use in training the \gls{lstm} model. The trained model is loaded during environment initialization and is used to predict a building's indoor dry-bulb temperature given a sequence of previously observed temperatures, loads, weather and temporal variables. With the provision of OpenStudio models in the \gls{eulp} dataset \cite{wilson_end-use_2022} that are convertible to EnergyPlus models as well as the work in \cite{bs2023_1404} to apply the \gls{eulp} dataset to create CityLearn input datasets from it, temperature dynamics models can be developed for a plethora of building configurations and locations across the \gls{usa} for use in control benchmarking in CityLearn.

CityLearn also provides the functionality to assess the resiliency of control algorithms during power outage events. During a power outage event, the grid is unable to provide buildings with electricity and agents can only make use of the flexibility provided by \glspl{der} in buildings to satisfy loads otherwise, risk thermal discomfort and unserved energy during the event. Whereas, during normal operation, there is unlimited supply from the grid. The outage signals are either statically defined as a time series or generated using some stochastic model. We provide a stochastic power outage model based on \gls{saifi} and \gls{caidi} distribution system reliability metrics \cite{IEEEGuideElectric2012} such that it is customizable and adaptable by providing location-specific metrics. The stochastic signals are generated by sampling $n$ instances from a binomial distribution to select days that experience power outage where the probability of a day having an outage, $p$, is the ratio of \gls{saifi} to number of days in a year (365). The start time step index for the outage on each day is then randomly selected from a uniform distribution. Finally, the duration of each power outage event is set by sampling from an exponential distribution with a scale set to \gls{caidi}.

Being an open-source environment, CityLearn can be customized to consider models of other \glspl{der} beyond what is contained in the as-provided code base including, \glspl{ev}, solar collector, and other controllable plug loads. It also comes prepackaged with a number of environment configuration datasets that have been used in The CityLearn Challenge\deleted[id=del]{ editions} \cite{nweye_citylearn_2022} and its code base\footnote{\url{https://www.citylearn.net}} has been extensively documented to support its application.
\section{Application} \label{sec:simulation}
In the following subsections, we describe the dataset used to generate the environment for each control simulation example as well as descriptions of the different environment, objective, and control algorithm combinations that make up the simulations. The source code for our work is made available on GitHub\footnote{\url{https://github.com/intelligent-environments-lab/simbuild-2024-citylearn/tree/v0.0.1}}.

\subsection{Dataset}
We use a dataset of two synthetic residential single-family buildings, B1 and B2, in Austin, Texas, \gls{usa} that are summarized in \cref{tab:building_metadata}. These buildings are sampled from building energy models in \cite{wilson_end-use_2022} that are representative of the residential and commercial building in the \gls{usa}. Both buildings have the same geometry and floor area but different load profile, baseline monthly consumption, construction decade, and \gls{bes} sizing. Each building has a \gls{dhw} \gls{tes} and \gls{bess}-\gls{pv} system for load shifting as well as controllable heat pump for load shedding. The environment is based on a one-month summer period of June 2018 thus, the heat pump is used only for cooling load satisfaction. An electric heater is used to satisfy \gls{dhw} heating load and charge the \gls{dhw} \gls{tes}. \added{The \gls{dhw} \gls{tes} has equal maximum charge and discharge rates equivalent to the electric heater's nominal power. }The \gls{dhw} \gls{tes} capacity is conservatively sized to equal half the maximum hourly \gls{dhw} heating demand in the one-month period, whereas the heat pump and electric heater are sized to satisfy the maximum hourly cooling and \gls{dhw} heating loads, respectively. The \gls{bess} is the equivalent of a real-world manufacturer's specifications with\added{ equal maximum charge and discharge nominal power of 3.3kW for B1 and 1.6kW for B2, and} a 20\% depth-of-discharge\deleted[id=del]{, while t}\added{. T}he \gls{pv} system is conservatively sized for a 21.5\% \gls{zne} scenario.

\begin{table}[!htb]
    \centering
    \caption{Environment building metadata. The average daily profile in the train period for cases with \gls{pv} generation (solid line) and without \gls{pv} generation (dotted line) are shown as well as the one-month electricity consumption without control and \gls{pv} generation.}
    \label{tab:building_metadata}
    \begin{tabular}{lrr}
        \hline
        \deleted[id=del]{\bf Building ID} & \bf B1 & \bf B2 \\
        \hline
        Geometry & {\includegraphics[height=0.55in]{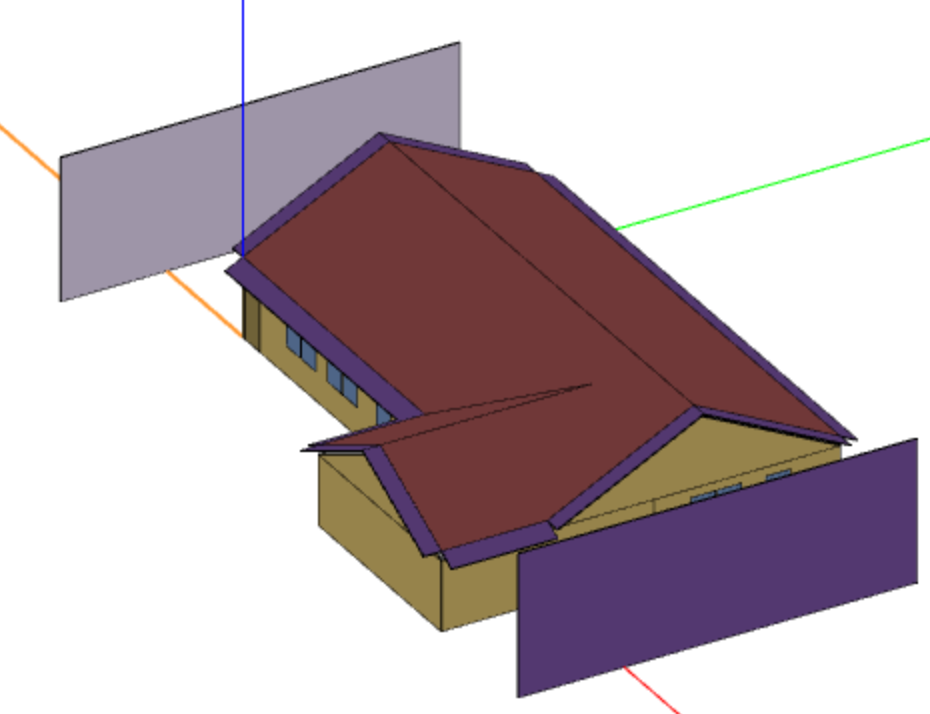}} & {\includegraphics[height=0.55in]{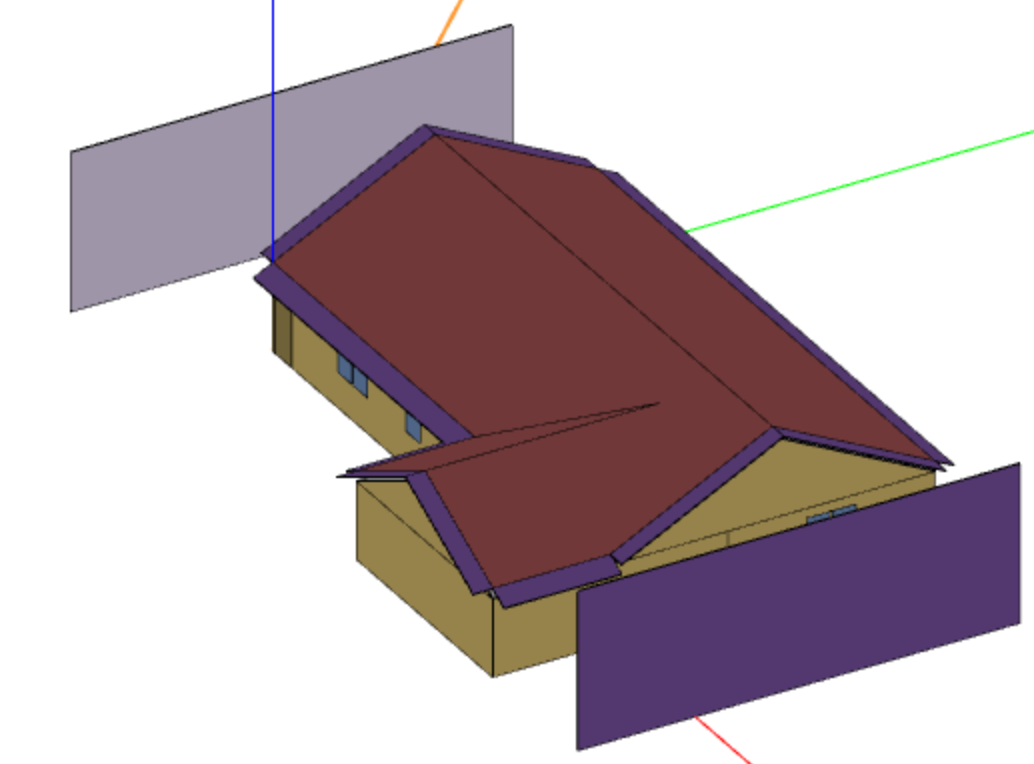}} \\
        Avg. daily profile & {\includegraphics[height=0.26in]{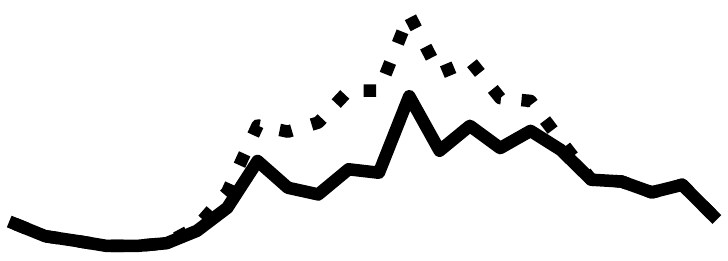}} & {\includegraphics[height=0.26in]{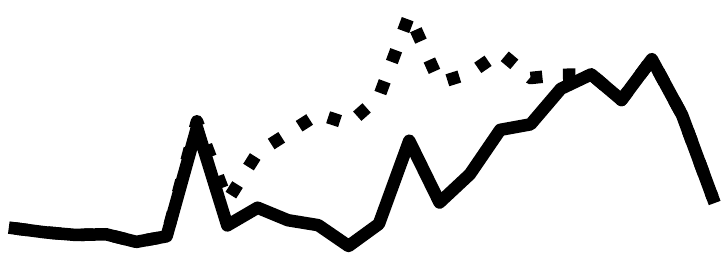}} \\
        Consumption (kWh) & 791.3 & 1293.8 \\
        Vintage & 1980s & 1970s \\
        Floor area (m\textsuperscript{2}) & 157.0 & 157.0 \\
        Heat pump (kW) & 2.3 & 2.8 \\
        \gls{dhw} \deleted[id=del]{H}\added{h}eater (kW) & 3.7 & 6.3 \\
        \gls{dhw} \gls{tes} (kWh) & 1.7 & 2.8 \\
        \gls{bess} (kWh) & 4.0 & 3.3 \\
        \added{\gls{bess} (kW)} & \added{3.3} & \added{1.6} \\
        \gls{pv} (kW) & 1.2 & 2.4 \\
        \hline
    \end{tabular}
\end{table}

Other supplementary datasets are 2018 weather data from the Austin Camp Mabry weather station that is provided by \cite{wilson_end-use_2022}, Austin Energy \gls{tou} rates in \cref{tab:tou_rate} that is reflective of its 2018 \gls{tou} pilot program \cite{city_of_austin_city_2017} and excludes any other static base rates. We also leverage the \gls{ercot} fuel mix dataset \cite{electric_reliability_council_of_texas_fuel_2021} and reported \gls{ghg} emissions for different energy sources in \cite{moomaw2011} to calculate hourly grid CO\textsubscript{2}e intensity (kgCO\textsubscript{2}e/kWh).

\begin{table}[!htb]
    \centering
    \caption{\Gls{tou} electricity rate.}
    \label{tab:tou_rate}
    \begin{tabular}{lrr}
        \hline
        & \multicolumn{2}{c}{\bf Rate (\$/kWh)} \\
        \bf Time & \bf Weekday & \bf Weekend \\
        \hline
        7 AM - 3 PM (Mid-Peak) & 0.0291 & 0.0289 \\
        3 PM - 6 PM (On-Peak) & 0.0587 & 0.0289 \\
        6 PM - 10 PM (Mid-Peak) & 0.0291 & 0.0289 \\
        10 PM - 7 AM (Off-Peak) & 0.0289 & 0.0289 \\
        \hline
    \end{tabular}
\end{table}

Using the framework in \cite{bs2023_1404}, we convert the building and supplementary datasets to a CityLearn input dataset. \added{It takes five minutes on average for us to train a single building's \gls{lstm} temperature dynamics model on a 16GB RAM, Intel Core i7 1255U CPU machine. }We use the initial 13 days in the one-\added{month }period for control agent hyperparameter tuning and training (train period) while the remaining 17 days are used for inference and evaluation of the control objectives (test period).

\subsection{Control Simulations}
We demonstrate the use of either \deleted[id=del]{a }\gls{rbc} or \gls{rlc} \deleted[id=del]{agent }to manage \deleted[id=del]{\gls{ess}}\added{the \glspl{ess}} and heat pump\deleted[id=del]{ power control}\added{ in the buildings}. \added{The \gls{bess} and \gls{dhw} \gls{tes} control actions received from the agent are the proportions of their capacities to be charged or discharged $\in [-1, 1]$ whereas, that of the heat pump is the proportion of its nominal power $\in [0, 1]$.}\deleted[id=del]{ towards minimizing} 

\added{The simulations are designed to minimize at least, }one of four building-level objectives and one district-level objective (\cref{tab:objective}), including cost (\cref{eqn:cost_objective}), emissions (\cref{eqn:emissions_objective}), discomfort (\cref{eqn:discomfort_objective}), consumption (\cref{eqn:consumption_objective}), and average daily peak (\cref{eqn:average_daily_peak_objective}). In the aforementioned equations, $t$ is the simulation time step, $n$ is the total number of time steps, $d$ is the day-of-year index, $h$ is the number of time steps in a day, which in our case is $h=24$. $e$ is the building-level electricity consumption in kWh, $P$ is district-level electricity demand in kW, $R$ is the electricity rate in \$/kWh, $G$ is the CO\textsubscript{2}e intensity in kgCO\textsubscript{2}e/kWh, $T$ is the indoor dry-bulb temperature in $^\circ$C, and $T_{\textrm{spt}}$ is the indoor dry-bulb temperature setpoint. 

\begin{table}[!htb]
    \centering
    \caption{Control objectives to minimize}
    \label{tab:objective}
    \begin{tabularx}\columnwidth{@{}lML@{}}
        \hline
        \bf Objective & \multicolumn{1}{l}{\bf Definition} & \multicolumn{1}{r}{} \\
        \hline
        Cost & \sum_{t=0}^{n-1}\textrm{max}(e(t), 0) \times R(t) & eqn:cost_objective \\
        Emissions & \sum_{t=0}^{n-1}\textrm{max}(e(t), 0) \times G(t) & eqn:emissions_objective \\
        Discomfort & \sum_{t=0}^{n-1}\lvert T(t) - T_{\textrm{spt}}(t) \rvert & eqn:discomfort_objective \\
        Consumption & \sum_{t=0}^{n-1}\textrm{max}(e(t), 0) & eqn:consumption_objective \\
        Avg. Daily Peak & \frac{\Big(\sum_{d=0}^{(n \div h) - 1} \textrm{max}(P(d))\Big) \cdot h}{n} & eqn:average_daily_peak_objective \\
        \hline
    \end{tabularx}
\end{table}

There is a total of 17 environment, objective, and control algorithm combinations we evaluate that vary in the type of algorithm complexity, reward function for \gls{rlc} algorithms, control objective, and controlled \glspl{der} including \gls{dhw} \gls{tes}, \gls{bess}-\gls{pv} system, and heat pump as summarized in \cref{tab:simulation_matrix}.

\begin{table*}[!htb]
    \centering
    \footnotesize
    \caption{Environment, objective, and control algorithm combinations that vary in the type of control algorithm, reward function for \gls{rlc} algorithms, control objective, and controlled \glspl{der}.}
    \label{tab:simulation_matrix}
    \begin{tabular}{lrlllcccc}
        \hline
        \bf Configuration ID & \bf Bldgs. & \bf Objective & \bf Reward & \bf Control Algo. & \bf \gls{dhw} & \bf \gls{bess} & \bf \gls{pv} & \bf Heat Pump \\
        \hline
        \textit{x-b1\_b2-x-x} & 2 & No Control & - & - & & & & \\
        \textit{x-b1\_b2-x-pv} & 2 & No Control & - & - & & & \ding{51} & \\
        \textit{rbc-b1-c-dhw} & 1 & Cost & - &  Cost-\gls{rbc} & \ding{51} & & & \\
        \textit{rbc-b1-e-dhw} & 1 & Emissions & - & Emission-\gls{rbc} & \ding{51} & & & \\
        \textit{rbc-b1-c-bess\_pv} & 1 & Cost & - & Cost-\gls{rbc} & & \ding{51} & \ding{51} & \\
        \textit{rbc-b1-e-bess\_pv} & 1 & Emissions & - & Emission-\gls{rbc} & & \ding{51} & \ding{51} & \\
        \textit{rbc-b1-c-dhw\_bess\_pv} & 1 & Cost & - & Cost-\gls{rbc} & \ding{51} & \ding{51} & \ding{51} & \\
        \textit{rbc-b1-e-dhw\_bess\_pv} & 1 & Emissions & - & Emission-\gls{rbc} & \ding{51} & \ding{51} & \ding{51} & \\
        \textit{rbc-b1\_b2-p-bess\_pv} & 2 & Peak & - & Peak-\gls{rbc} & & \ding{51} & \ding{51} & \\
        \textit{rlc-b1-c-dhw} & 1 & Cost & \cref{eqn:cost_reward_function} &  \acrshort{sac}-\gls{rlc} & \ding{51} & & & \\
        \textit{rlc-b1-e-dhw} & 1 & Emissions & \cref{eqn:emissions_reward_function} & \acrshort{sac}-\gls{rlc} & \ding{51} & & & \\
        \textit{rlc-b1-c-bess\_pv} & 1 & Cost & \cref{eqn:cost_reward_function} &  \acrshort{sac}-\gls{rlc} & & \ding{51} & \ding{51} & \\
        \textit{rlc-b1-e-bess\_pv} & 1 & Emissions & \cref{eqn:emissions_reward_function} & \acrshort{sac}-\gls{rlc} & & \ding{51} & \ding{51} & \\
        \textit{rlc-b1-c-dhw\_bess\_pv} & 1 & Cost & \cref{eqn:cost_reward_function} &  \acrshort{sac}-\gls{rlc} & \ding{51} & \ding{51} & \ding{51} & \\
        \textit{rlc-b1-e-dhw\_bess\_pv} & 1 & Emissions & \cref{eqn:emissions_reward_function} & \acrshort{sac}-\gls{rlc} & \ding{51} & \ding{51} & \ding{51} & \\
        \textit{rlc-b1-d\_o-hp} & 1 & Discomfort \& Consumption & \cref{eqn:discomfort_and_consumption_reward_function} & \acrshort{sac}-\gls{rlc} & & & & \ding{51} \\
        \textit{rlc-b1\_b2-p-bess\_pv} & 2 & Peak & \cref{eqn:average_daily_peak_reward_function} & \acrshort{sac}-\gls{rlc} & & \ding{51} & \ding{51} & \\
        \hline
    \end{tabular}
\end{table*}

\subsubsection{Baseline (No Control)}
There are two baseline configurations, \textit{x-b1\_b2-x-x} and \textit{x-b1\_b2-x-pv} that include both B1 and B2 in the environment where there are no controlled \glspl{der}. However, \textit{x-b1\_b2-x-pv} includes a \gls{pv} system in each building to augment electricity provision from the grid.

\subsubsection{Rule-Based Control}
We define seven \gls{rbc}-based configurations. \textit{rbc-b1-c-dhw} and \textit{rbc-b1-e-dhw} include only building B1 equipped with \gls{dhw} \gls{tes} control but differ in their control objectives, with the former being cost reduction and the latter being emissions reduction. Accordingly, the \gls{rbc} agents are tailored to meet these objectives, where we tune a cost-reduction-based \gls{rbc} according to the rates in \cref{tab:tou_rate}. On weekdays, the \gls{rbc} is set to charge the \gls{ess} to capacity during the off-peak hours of 10 PM to 7 AM, discharge 50\% of its capacity during the on-peak period of 3 PM to 6 PM and discharge the remaining energy at other times. During the weekend, when a cheap flat rate is used, the \gls{ess} is charged at a constant rate.

We tune the emission-reduction-based \gls{rbc} by qualitatively analyzing the ERCOT average daily CO\textsubscript{2}e intensity profile during the train period. The grid CO\textsubscript{2}e intensity is the least prior to 8 AM, peaks around 1 PM but sustains a high intensity between noon and 11 PM. Thus, the emission-reduction-based \gls{rbc} is set to fully charge the \gls{ess} at a constant rate prior to 8 AM and discharge the stored energy at a constant rate between 12 PM and 11 PM.

The \textit{rbc-b1-c-bess\_pv} and \textit{rbc-b1-e-bess\_pv} configurations replace the \gls{dhw} \gls{tes} with a \gls{bess}-\gls{pv} system whereas, the \textit{rbc-b1-c-dhw\_bess\_pv}, and \textit{rbc-b1-e-dhw\_bess\_pv} augment the \gls{dhw} \gls{tes} with the \gls{bess}-\gls{pv} system. The \textit{rbc-b1\_b2-p-bess\_pv} configuration uses both buildings B1 and B2 with a district-level objective of average daily peak reduction (\cref{eqn:average_daily_peak_objective}) and makes use of a peak-reduction-based \gls{rbc} to control the \gls{bess}. We tune the \gls{rbc} by inferring the peak hours in each building's daily profile shown in \cref{tab:building_metadata} such that it charges the \gls{ess} to capacity before 6 AM at steady and then, discharges it at a steady rate between 6 AM and 11 PM.

\subsubsection{Reinforcement Learning Control}
We design eight \gls{rlc} configurations, of which seven are the \gls{rlc} equivalent of the seven \gls{rlc} configurations. We make use of the \gls{sac} \gls{rlc} algorithm \cite{haarnoja_soft_2018} and take advantage of its default implementation in the Stable-Baselines3 Python package \cite{raffin_stable-baselines3_2021}. Each configuration environment is used to train an agent on 150 epochs of the 13-day train period.

We define a reward function in \cref{tab:reward_function} for each \gls{rlc} objective in \cref{tab:simulation_matrix}. The \textit{rlc-b1-c-dhw}, \textit{rlc-b1-c-bess\_pv}, and \textit{rlc-b1-c-dhw\_bess\_pv} configurations have the same cost objective and reward function (\cref{eqn:cost_reward_function}). \Cref{eqn:cost_reward_function} is similar to the objective function defined in, \cref{eqn:cost_objective} with the exception that it only considers the cost at the current time step. Given that the goal in training an \gls{rlc} agent is to maximize the cumulative reward, we apply a negative sign to the calculated reward so that large costs have lower rewards. A similar approach is taken for the reward function used in the \textit{rlc-b1-e-dhw}, \textit{rlc-b1-e-bess\_pv}, and \textit{rlc-b1-e-dhw\_bess\_pv} configurations where the emission reward function (\cref{eqn:emissions_reward_function}) is the negative equivalent of \cref{eqn:emissions_objective} for a single time step.

\begin{table}[!htb]
    \centering
    \footnotesize
    \caption{Reward functions.}
    \label{tab:reward_function}
    \begin{tabularx}\columnwidth{@{}lML@{}}
        \hline
        \bf Penalty & \multicolumn{1}{l}{\bf Definition} & \multicolumn{1}{r}{} \\
        \hline
        Cost & -\textrm{max}(e(t), 0) \times R(t) & eqn:cost_reward_function \\
        Emissions & -\textrm{max}(e(t), 0) \times G(t) & eqn:emissions_reward_function \\
        Discomfort \& Cons. & \begin{cases} -m \times \lvert T(t) - T_{\textrm{spt}}(t) \rvert, \ \textrm{if} \ T(t) < T_{\textrm{spt}}(t)\\ -\lvert T(t) - T_{\textrm{spt}}(t) \rvert, \ \textrm{otherwise}\end{cases} & eqn:discomfort_and_consumption_reward_function \\
        Avg. Daily Peak & -\textrm{max}(P(t), 0) & eqn:average_daily_peak_reward_function \\
        \hline
    \end{tabularx}
\end{table}

The \textit{rlc-b1-d\_o-hp} configuration is multi-objective where, the goal is to minimize electricity consumption (\cref{eqn:consumption_objective}) while maintaining comfort (\cref{eqn:discomfort_objective}). We encode both objectives in the discomfort and consumption reward function defined in \cref{eqn:discomfort_and_consumption_reward_function} where the penalized delta between $T$ and $T_{\textrm{spt}}(t)$ is scaled up by a factor of $m$, when $T < T_{\textrm{spt}}(t)$ to prevent the \gls{rlc} agent from over-cooling the building which has an adverse effect of increased electricity consumption.

The \textit{rlc-b1\_b2-p-bess\_pv} configuration only differs from \textit{rbc-b1\_b2-p-bess\_pv} in control algorithm. \textit{rlc-b1\_b2-p-bess\_pv} utilizes the \gls{sac}-gls{rlc} agent and average daily peak reward function (\cref{eqn:average_daily_peak_reward_function}) to minimize average daily peak (\cref{eqn:average_daily_peak_objective}) where the reward function is set to penalize high electricity demand which by proxy penalizes peaks.

We set each \gls{rlc}-based configuration to use a subset of 12 observations. The hour and day-of-week observations are used in all simulations. Net electricity consumption is excluded in the configurations that include heat pump control and discomfort objective. The electricity rate and CO\textsubscript{2}e intensity observations are only relevant to configurations that seek to minimize cost and emissions, respectively. Solar generation and \gls{bess} \gls{soc} observations are included in environments that have \gls{bess}-\gls{pv} systems in the buildings whereas \gls{dhw} \gls{tes} \gls{soc} is observed only in environments that have \gls{dhw} \gls{tes}. Outdoor and indoor dry-bulb temperature, indoor dry-bulb temperature setpoint and the absolute delta between indoor dry-bulb temperature and indoor dry-bulb temperature setpoint are active only in environments with heat pump control.
\section{Results} \label{sec:results}
\subsection{Single Building Cost or Emissions Objective}
\begin{figure}[!htb]
    \centering
    \begin{subfigure}[]{\columnwidth}
        \centering
        \includegraphics[width=\columnwidth]{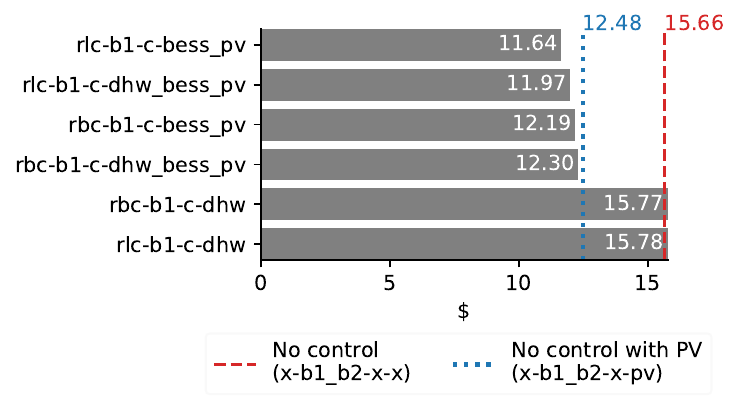}
        \caption{Cost (\$).}
        \label{fig:images/cost_comparison}
    \end{subfigure}
    \begin{subfigure}[]{\columnwidth}
        \centering
        \includegraphics[width=\columnwidth]{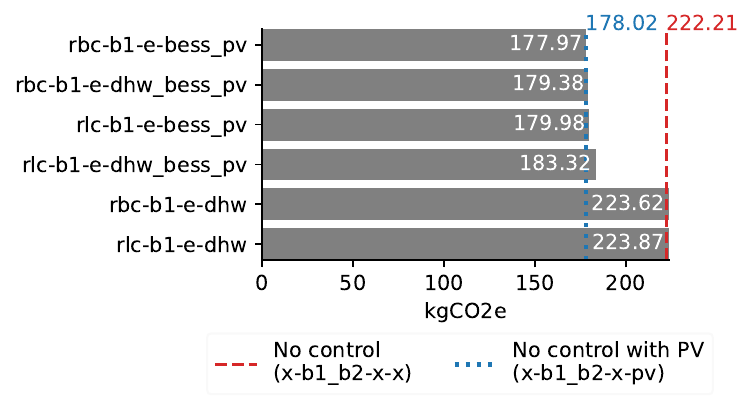}
        \caption{Emissions (kgCO\textsubscript{2}e).}
        \label{fig:images/emission_comparison}
    \end{subfigure}
    \caption{Cost (\$) and emissions (kgCO\textsubscript{2}e) from either \gls{rbc} or \gls{rlc} of \gls{dhw} \gls{tes}, \gls{bess}-\gls{pv} system or both when the control objective is cost or emission. The dashed red line shows the cost for a no-control scenario and the dotted blue line shows the cost for a no-control scenario but with solar generation to augment electricity from the grid.}
    \label{fig:images/cost_and_emission_comparison}
\end{figure}

\Cref{fig:images/cost_and_emission_comparison} shows cost (\$) and emissions (kgCO\textsubscript{2}e) in building B1 from either \gls{rbc} or \gls{rlc} of \gls{dhw} \gls{tes}, \gls{bess}-\gls{pv} system or both when the control objective is cost (\cref{fig:images/cost_comparison}) or emission (\cref{fig:images/emission_comparison}), compared to baseline configurations with no control (\textit{x-b1\_b2-x-x}) and no control but solar generation (\textit{x-b1\_b2-x-pv}) in the 17-day test period. We show that the \gls{pv} system advantage in terms of cost and emissions reduction is approximately 20.0\%, similar to the 21.5\% \gls{zne} that the \gls{pv} system was sized for. There is no cost or emissions reduction benefit from \gls{dhw} \gls{tes} without \gls{pv} system control when either an \gls{rbc} or \gls{rlc} agent is used. The largest cost savings compared to \textit{x-b1\_b2-x-pv} is \$0.83 (6.7\%) that is provided by the \gls{rl}-controlled \gls{bess}-\gls{pv} system (\textit{rlc-b1-c-bess\_pv}). Including a \gls{dhw}-\gls{tes} to the \gls{bess}-\gls{pv} environment (\textit{rlc-b1-c-dhw\_bess\_pv}) slightly increases the cost but still performs better than the rule-base-controlled \gls{bess}-\gls{pv} system. Both \gls{rbc} configurations that include solar generation in the building have higher costs than their \gls{rlc} alternative but still perform better than the baseline in magnitudes of 2.3\% and 1.4\% for \textit{rbc-b1-c-bess\_pv} and \textit{rbc-b1-c-dhw\_bess\_pv}.

In contrast and shown in \cref{fig:images/emission_comparison}, neither \gls{rbc} nor \gls{rlc} provides emissions reduction irrespective of the available \glspl{ess} when paired with a \gls{pv} system and our manually tuned \gls{rbc} performs marginally better than the advanced \gls{rlc} agent for an emissions reduction objective. The \textit{rlc-b1-e-dhw\_bess\_pv} environment in fact, increases emissions by approximately 3.0\%.

\begin{figure*}[!htb]
    \centering
    \begin{subfigure}[]{\textwidth}
        \centering
        \includegraphics[width=\textwidth]{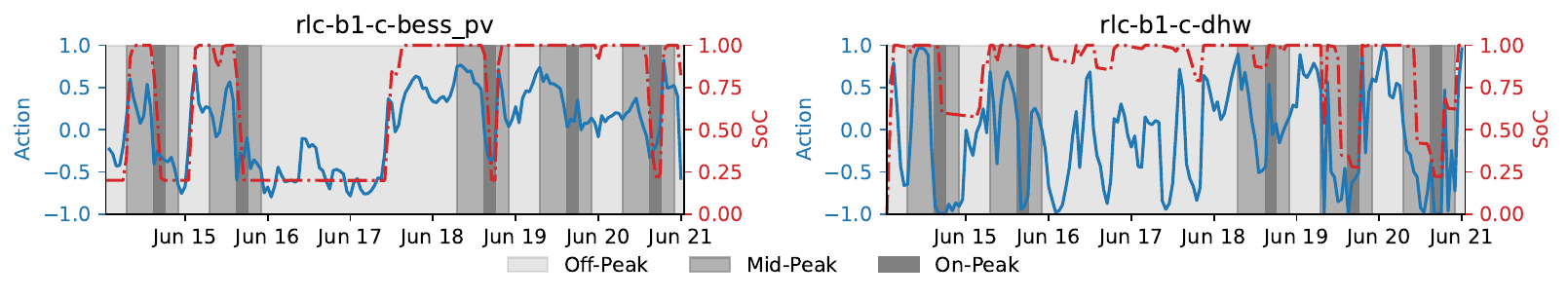}
        \caption{\gls{bess}-\gls{pv} system (right) and \gls{dhw} \gls{tes} (left) control with cost objective.}
        \label{fig:cost_soc_action_time_series}
    \end{subfigure}
    \begin{subfigure}[]{\textwidth}
        \centering
        \includegraphics[width=\textwidth]{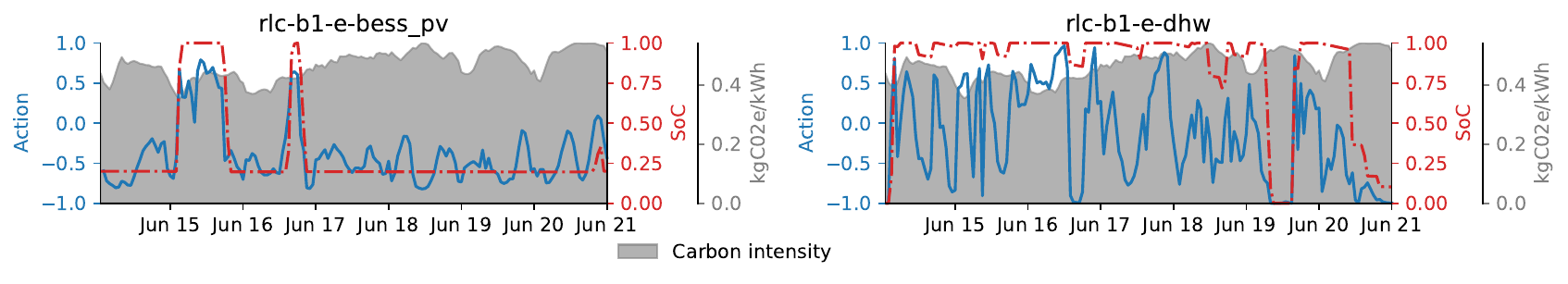}
        \caption{\gls{bess}-\gls{pv} system (right) and \gls{dhw} \gls{tes} (left) control with emissions objective.}
        \label{fig:emission_soc_action_time_series}
    \end{subfigure}
    \caption{\gls{rlc} action and consequent \gls{ess} \gls{soc} trend in the initial seven days of the two-week evaluation period when the control objective is to either minimize cost or emissions.}
    \label{fig:soc_action_time_series}
\end{figure*}

\Cref{fig:soc_action_time_series} shows the \gls{rlc} actions in the initial seven days of the test period for the configurations that have one \gls{ess} with cost reduction objective (\textit{rlc-b1-c-dhw} and \textit{rlc-b1-c-dhw\_bess\_pv}), or emissions reduction objective (\textit{rlc-b1-e-dhw} and \textit{rlc-b1-e-dhw\_bess\_pv}). The resulting changes in the \gls{ess} \gls{soc} as a consequence of these actions are also shown as well as shaded regions to indicate the \gls{tou} as defined in \cref{tab:tou_rate} as well as the CO\textsubscript{2}e intensity. The \gls{soc} trends in \cref{fig:cost_soc_action_time_series} show that with the cost reward function (\cref{eqn:cost_reward_function}), the \gls{rlc} agent learns to charge the \glspl{ess} during the off-peak period while discharging mainly during the on-peak period. However, the charge and discharge rate are rapid, reaching full charge and discharge depth within few time steps. Although, the agent calls for charging during the mid-peak period, this \gls{tou} coincides with when the \gls{ess} is charged to capacity i.e., $\textrm{\gls{soc}}=1$ thus, has no effect on the \gls{soc}. There is also an unusual call to discharge throughout June 16, a Saturday, when electricity is cheapest. Compared to the \gls{bess}, the \gls{dhw} \gls{tes} rarely discharges its full capacity as the agent calls for charge action, even when the \gls{soc} is well above 75.0\%. There is also no change in \gls{dhw} \gls{tes} \gls{soc} in some instances where the agent calls for energy discharge.

The results for an emissions reduction objective, depicted in \cref{fig:emission_soc_action_time_series}, show persistent call for discharge of the \gls{bess} by the \gls{rlc} agent for almost four consecutive days even though the \gls{bess} has reached its depth-of-discharge. The agent does take advantage of the lower CO\textsubscript{2}e intensity to charge however, there is a very mild variation in carbon intensity for the evaluated period as there are no distinct periods of significantly low emissions to take advantage of. In contrast, the \gls{rlc} agent controlling the \gls{dhw} \gls{tes} shows no clear pattern in selected actions as it charges and discharges at random. We also observe a similar \gls{soc} trend as with the cost objective where the \gls{tes} rarely depletes below 75.0\% \gls{soc}.

\subsection{Single Building Discomfort and Consumption Objective}
In \cref{tab:discomfort_consumption_summary}, we show the effect of varying the multiplier, $m$ in the discomfort and consumption reward function on discomfort, average discomfort when over-cooling ($\overline{T_{\Delta}}<0$) and under-cooling ($\overline{T_{\Delta}}>0$), and consumption objectives when used in heat pump control environment,  \textit{rlc-b1-d\_o-hp}. The percent changes in objective by setting $m=3$, $m=6$, and $m=12$ are compared to the baseline case, $m=1$, where there is no penalty for over-cooling i.e., objective is reduced to discomfort reduction alone. We see that by setting $m=3$ to activate the multi-objective reward, we reduce consumption by 3.0\% but also discomfort by 3.0\% through 50.0\% decrease in over-cooling. In contrast, $m=3$ increases under-cooling by only 20.0\%. Larger increments in $m$ to $m=6$ and $m=12$ show an increase in discomfort of almost 100.0\% compared to the baseline but up to 12.0\% decrease in consumption. However, the average temperature difference between the indoor dry-bulb temperature and setpoint for the selected $m$ values, irrespective of over-cooling or under-cooling, is less than 1.5C.

\begin{table*}
    \centering
    \caption{Effect of varying $m$ in discomfort and consumption reward function, \cref{eqn:average_daily_peak_reward_function}, on discomfort, average discomfort when over-cooling ($\overline{T_{\Delta}}<0$) and under-cooling ($\overline{T_{\Delta}}>0$), and consumption objectives when used in heat pump control environment,  \textit{rlc-b1-d\_o-hp}. The percent change in objective value, by varying $m$ is compared to the baseline case, $m=1$, where there is no penalty for increased consumption from over-cooling. Improvement in an objective is highlighted in \textcolor{blue}{blue} while deterioration is highlighted in \textcolor{red}{red}.}
    \label{tab:discomfort_consumption_summary}
    \begin{tabular}{rrrrr}
    \hline
         & \bf Discomfort & \bf Over-cool ($\overline{\bm T_{\bm \Delta}}\bm < \bm 0$) & \bf Under-cool ($\overline{\bm T_{\bm \Delta}}\bm > \bm 0$) & \bf Consumption \\
          $\bm m$ & \bf ($^\circ$Ch) & \bf ($^\circ$C) & \bf ($^\circ$C) & \bf (kWh) \\
         \hline
        1 & 201.8 & $0.4\pm0.5$ & $0.5\pm0.5$ & 384.8 \\
        3 & 195.7 (\textcolor{blue}{-3.0\%}) & $0.2\pm0.2$ (\textcolor{blue}{-50.0\%}) & $0.6\pm0.5$ (\textcolor{red}{20.0\%}) & 373.0 (\textcolor{blue}{-3.0\%}) \\
        6 & 315.9 (\textcolor{red}{56.5\%}) & $0.3\pm0.2$ (\textcolor{blue}{-25.0\%}) & $0.9\pm0.5$ (\textcolor{red}{80.0\%}) & 352.5 (\textcolor{blue}{-8.3\%}) \\
        12 & 399.6 (\textcolor{red}{98.0\%}) & $0.1\pm0.4$ (\textcolor{blue}{-75.0\%}) & $1.0\pm0.6$ (\textcolor{red}{100.0\%}) & 338.1 (\textcolor{blue}{-12.0\%}) \\
        \hline
    \end{tabular}
\end{table*}

\subsection{District Average Daily Peak Objective}
\begin{figure}[!htb]
    \centering
    \includegraphics[width=\columnwidth]{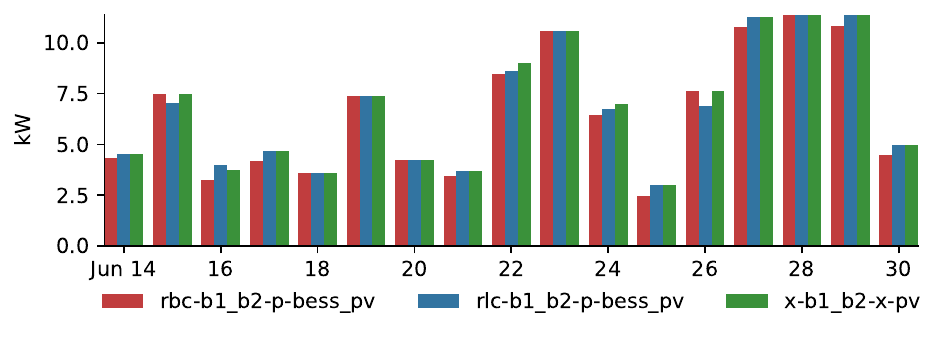}
    \caption{District-level daily peak load for two-building environments where each building has \gls{bess}-\gls{pv} system.}
    \label{fig:daily_peak}
\end{figure}

The daily district-level peak load is shown in \cref{fig:daily_peak} for the baseline no-control with \gls{pv} generation (\textit{x-b1\_b2-x-pv}) configuration and the two configurations with peak reduction objective but differing control algorithms: \textit{rbc-b1\_b2-p-bess\_pv} and \textit{rlc-b1\_b2-p-bess\_pv}. Compared to the baseline, the \gls{rbc} is able to reduce the peak on ten out of the 17-day test period at an average of 8.4\%. In contrast, the \gls{rlc} only reduces the peak below the baseline on four days at an average of 5.9\%. The \gls{rbc} never peaks higher than the baseline, in contrast to the \gls{rlc} that exceeds the baseline thrice. There are eleven days when the \gls{rbc} is able to perform better than the \gls{rlc} while the \gls{rlc} outperforms the \gls{rbc} on only two days. There are four days when neither control approach is able to reduce peak load compared to the baseline. In summary, the \gls{rbc} provides a 2.6\% advantage in average daily peak reduction over \gls{rlc}. However, neither agent is able to reduce the highest peak on June 28.

\section{Discussion} \label{sec:discussion}
Our results for the \textit{rbc-b1-c-dhw}, \textit{rlc-b1-c-dhw}, \textit{rbc-b1-e-dhw}, and \textit{rls-b1-e-dhw} configurations where we shift \gls{dhw} loads in B1 using an \gls{rbc} or \gls{rlc} agent in order to reduce cost and emissions show slightly higher costs and emissions compared to the baseline (\cref{fig:images/cost_and_emission_comparison}). We infer that this observation is as a result of our conservatively-sized \gls{dhw} \gls{tes} which is only half of the peak hourly demand. This sizing is a design choice that is characteristic of the dataset and its original curation for use in The CityLearn Challenge 2023\footnote{\url{https://www.aicrowd.com/challenges/neurips-2023-citylearn-challenge}} as it makes it challenging to leverage largely sized \glspl{bes} to reduce electricity consumption, costs or emissions. Moreover, the \gls{dhw} demand in B1 is only 7\% of the total building demand, and there is no \gls{dhw} demand during 330/407 control time steps. Our observation in \cref{fig:soc_action_time_series} where the \gls{tes} almost never depletes below 75.0\% capacity irrespective of discharge action by the \gls{rlc} agent is attributed to the infrequent occurrence of \gls{dhw} demand and can make association of actions to changes in \gls{soc} observations challenging for the agent. In the same vein, the choice or \gls{bess}-\gls{pv} size plays a role in the amount of energy flexibility it is able to provide.  In this work, we have used a \gls{pv} sized for 21.5\% \gls{zne} while the \gls{bess} capacity is able to meet almost 100.0\% of hourly demand. Previous work has however shown that for a single-family grid-interactive building designed for 100.0\% \gls{zne} in Fontana, California with similar warm climate as Austin, Texas, up to 27.0\% and 37.0\% annual cost and emissions reduction respectively achievable \cite{nweye_merlin_2023} when paired with an \gls{rl}-controlled \gls{bess}.

Although, the \gls{tou} rate we use in \cref{tab:tou_rate} is three-tiered, there is negligible difference in rates value across the tiers, making for a weak cost signal. Likewise, the variance in carbon intensity is less than 1.0\% making it also a weak signal. Typically, \gls{tou} rates are reflective of how clean the electricity grid is and the time of peak demand. The average daily share of renewable energy sources in the \gls{ercot} grid in June 2018 is estimated at only 20.1\% with 5.5\% standard deviation \cite{electric_reliability_council_of_texas_fuel_2021}. This highlights the importance of simultaneously decarbonizing the supply-side as demand-side end uses are electrified or risk an adverse effect of increased emissions. Although no observation forecasts have been used in our work for \gls{rlc}, providing the agent with cost and CO\textsubscript{2}e intensity forecast over a short horizon could improve learning and performance.

We observe in \cref{fig:soc_action_time_series} that the \gls{rlc} agents learn to charge when electricity is cheap but take redundant actions, including calling for charging when the \glspl{ess} is fully charged or discharging when completely depleted. However, these futile actions have no associated penalty. We consider a reward function that has a penalty for actions that are unable to change the state of the \gls{ess} could improve the action quality of the agent.

The work by June et al. on occupant-centric \deleted[id=del]{lighting and }thermostat control \cite{park_hvaclearn_2020} has shown reduced setpoint overrides and energy consumption when occupant comfort and preferences are considered in control decisions. It is estimated from field studies that a median of up to 20.0\% energy savings are accrued from \gls{occ} \cite{jung_human---loop_2019}. CityLearn provides a framework for such \gls{occ} by employing any of two methodology. In the first methodology, ideal loads that satisfy comfort are maintained while a control agent learns to improve energy efficiency through \glspl{ess} load shifting. The second methodology, as we have shown in this work with heat pump control in the CityLearn v2 environment, is where both comfort and energy efficiency can be learned by way of reward function design. For a cooling season, we show that by penalizing indoor temperature below the setpoint, an agent is able to learn that over-cooling had an adverse effect of increased consumption. This reward function style has been used in the literature by Pinto et al. where exponents on the temperature delta when outside a comfort band are used to teach the agent to prefer under-cooling over over-cooling \cite{pinto_data-driven_2021}.

Our results from average daily peak reduction show better performance from using an \gls{rbc} agent that has been finely tuned to target energy discharge during peak periods compared to an \gls{rlc} agent that tries to learn to reduce the peaks using the electricity consumption as a reward signal. In \cite{nweye_real-world_2022}, Nweye et al. show that an expert tuned \gls{rbc} agent is able to outperform \gls{rlc} algorithms however, it may be beneficial to take advantage of the expert rules to train the \gls{rlc} agent offline. This method of training from expert systems been used in simple gaming applications \cite{argerich_tutor4rl_2020} and complex HVAC control \cite{dey_reinforcement_2023}.

Our dataset is constructed from the End-Use Load Profiles (EULP) for the U.S. Building Stock dataset \cite{wilson_end-use_2022} where we follow the methodology in \cite{bs2023_1404} to create our CityLearn environment. The two buildings we use (\cref{tab:building_metadata}) are each reflective of about 242 actual single-family buildings in Austin, Texas and in comparison to 2020 Residential Energy Consumption Survey (RECS) \cite{eiaRecs2020}, the total floor area of our buildings are near the average residential building floor area of 161m\textsuperscript{2} in Texas. The estimated average customer electricity consumption reported by Austin Energy for June 2018 is 1,144 kWh \cite{austin_energy_residential_2023}. This estimate is similar to electricity consumption of 1293.8 kWh  in building B2 during our June data period however, building B1 has a much lower consumption that is about 30\% less than the survey estimate. We attribute this disparity to the way heat pumps are modeled in CityLearn. The CityLearn heat pump model assumes a Carnot cycle with maximum theoretical efficiency that may not accurately reflect the heat pump operation and specification in the original \gls{eulp} building model. CityLearn is best applied to simple and quick comparative analysis amongst control algorithms at single-building or district levels with minimal design cost. Whereas, an environment that utilizes high-fidelity energy models e.g., BOPTEST \cite{blum_building_2021} for single-building control or DOPTEST \cite{arroyo_prototyping_2023} for district-control is better suited for analyses that are purposed for scrutiny of absolute energy and cost values. However, the challenge with using these high-fidelity environments is the limited system and building configurations users are restricted to such that implementing buildings from the \gls{eulp} dataset for large scale building stock studies will incur significant design costs.

Despite its potential, there are still open questions regarding the plug-and-play capabilities, performance, safety of operation, and learning speed of \gls{rlc}. The CityLearn Challenge is an opportunity to compete in investigating the potential of \gls{ai} and distributed control systems to tackle multiple problems within the built-environment domain. Other competitions in the energy domain focus on building load predictions \cite{miller_ashrae_2020}, grid power flow optimization \cite{holzer_grid_2021} and pathways to building electrification and decarbonization \cite{nyserda_nyserda_2022}. While these competitions provided a plethora of solutions that are pertinent to energy supply and demand-side management, The CityLearn Challenge is one of its kind with a focus on \gls{bes} control and control algorithm benchmarking for demand response \cite{nweye_citylearn_2022}. Previous editions of The CityLearn Challenge have investigated transferability of control policies,\deleted[id=del]{ and} a realistic implementation of a model-free \gls{rl} in buildings where training evaluation is done on a single four-year long episode, and \gls{bess} control in a real-world \gls{zne} community for energy flexibility maximization. Thus, CityLearn attracts a multidisciplinary participation audience including researchers, industry experts, sustainability enthusiasts and \gls{ai} hobbyists as an avenue for crowdsourcing solutions to building control problems using \gls{rl} or any other algorithm of choice.
\section{Conclusion} \label{sec:conclusion}
Our work here has shown 17 different building control applications in CityLearn. We show that simple \acrlong{rbc} and complex \acrlong{rlc} agents managing different \acrlong{der} can be benchmarked in a two-building residential district, based on realistic building stock energy models, towards optimizing either singular objective or multi-objective \acrlongpl{kpi}.

\added{ For our dataset, the results show that \gls{rlc} provides further cost reduction compared to \gls{rbc} while, neither algorithm provides substantial reduction in emissions irrespective of what \glspl{der} are available. Also \gls{rlc} can learn to reduce discomfort, overcooling and consequently consumption when reward function parameters are adequately tuned. Lastly, for our peak reduction objective, \gls{rbc} provides a 2.6\% advantage over \gls{rlc}.} 

The applications presented here are by no means an exhaustive demonstration of the CityLearn environment's functionalities, especially its use for control resiliency benchmarking. Instead, we envision that our contribution provides environment and control examples as well as their implementation source code for members of the built environment community whom are interested in using CityLearn to solve their \gls{bes} control problems.
% --------------------------------------------------------

\bibliographystyle{achicago}
\bibliography{references}

\end{document}